\documentclass[11pt, oneside]{article}   	% use "amsart" instead of "article" for AMSLaTeX format
\usepackage[margin=1in]{geometry}         % See geometry.pdf to learn the layout options. There are lots.
\usepackage{graphicx}				% Use pdf, png, jpg, or eps§ with pdflatex; use eps in DVI mode
\usepackage{amsmath}	
\usepackage{amssymb}
\usepackage{breqn}
\usepackage{natbib}
\usepackage{hyperref}
\usepackage{amsthm}
\usepackage{bm}
\usepackage{authblk}
\usepackage{listings}

\newcommand{\RN}[1]{%
  \textup{\uppercase\expandafter{\romannumeral#1}}%
}

\newtheorem{mydef}{Definition}

\newtheorem{myprop}{Proposition}
\newtheorem{mythm}{Theorem}

\hypersetup{
   bookmarks=true,         % show bookmarks bar?
   unicode=false,          % non-Latin characters in AcrobatÕs bookmark
   pdftoolbar=true,        % show AcrobatÕs toolbar?
   pdfmenubar=true,        % show AcrobatÕs menu?
   pdffitwindow=false,     % window fit to page when opened
   pdfstartview={FitH},    % fits the width of the page to the window
   pdftitle={My title},    % title
   pdfauthor={Author},     % author
   pdfsubject={Subject},   % subject of the document
   pdfcreator={Creator},   % creator of the document
   pdfproducer={Producer}, % producer of the document
   pdfkeywords={keyword1} {key2} {key3}, % list of keywords
   pdfnewwindow=true,      % links in new PDF window
   colorlinks=true,       % false: boxed links; true: colored links
   linkcolor=red,          % color of internal links (change box color with linkbordercolor)
   citecolor=blue,        % color of links to bibliography
   filecolor=magenta,      % color of file links
   urlcolor=cyan           % color of external links
}

\title{\textbf{A Note on Spherical Needlets}}
\author{Minjie Fan \thanks{Address for correspondence: Minjie Fan, Ph.D. candidate, Department of Statistics, University of California, Davis, One Shields Avenue, Davis, CA 95616, U.S.A. \\Email: {\tt mjfan@ucdavis.edu}}}
\affil{Department of Statistics, University of California, Davis}
\date{\today}							% Activate to display a given date or no date

\begin{document}
\maketitle
%\section{}
%\subsection{}
\begin{abstract}
Compared with the traditional spherical harmonics, the spherical needlets are a new generation of spherical wavelets that possess several attractive properties. Their double localization in both spatial and frequency domains empowers them to easily and sparsely represent functions with small spatial scale features. This paper is divided into two parts. First, it reviews the spherical harmonics and discusses their limitations in representing functions with small spatial scale features. To overcome the limitations, it introduces the spherical needlets and their attractive properties. In the second part of the paper, a Matlab package for the spherical needlets is presented. The properties of the spherical needlets are demonstrated by several examples using the package.
\end{abstract}
\section{Spherical Harmonics}
Plainly speaking, the spherical harmonics $\{Y_{lm}, l=0,1,\cdots, m=-l,\cdots,l\}$ are a natural extension of the Fourier basis functions to the domain of the unit sphere $\mathbb{S}^2$.  For convenience, we employ $(\theta, \phi), 0\leq \theta \leq \pi, 0 \leq \phi <2\pi$, where $\theta$ is the co-latitude and $\phi$ is the longitude, as the spherical coordinate of $\bm{x}\in \mathbb{S}^2$. The spherical harmonics share two crucial characteristics as the Fourier basis functions. First, they are eigenfunctions of the eigenvalue problem
\begin{equation}
\Delta_{\mathbb{S}^2} Y = -\lambda Y,
\end{equation}
where $\Delta_{\mathbb{S}^2}$ is the Laplace-Beltrami operator defined on $\mathbb{S}^2$, i.e.,
\begin{dmath}
\Delta_{\mathbb{S}^2}=\frac{1}{\sin \theta}\frac{\partial}{\partial \theta}\left(\sin \theta \frac{\partial}{\partial \theta}\right)+\frac{1}{\sin^2 \theta}\frac{\partial^2}{\partial \phi^2}.
\end{dmath}
Besides, they constitute an orthonormal basis of the Hilbert space $L^2(\mathbb{S}^2)$. Thus, for any function $T$ in $L^2(\mathbb{S}^2)$, it can be uniquely expanded as 
\begin{dmath}
T(\bm{x})=\sum \limits_{l=0}^{\infty}\sum \limits_{m=-l}^{l} a_{lm}Y_{lm}(\bm{x}),
\end{dmath}
where 
\begin{dmath}\label{alm}
a_{lm}=\int_{\mathbb{S}^2} T(\bm{x})\overline{Y}_{lm}(\bm{x})d\bm{x},
\end{dmath}
which are called spherical harmonic coefficients.

The spherical harmonics can be represented in terms of the associated Legendre polynomials.
\begin{mydef}\label{SH}
For any $l=0,1,\cdots, m=-l,\cdots,l$, the spherical harmonics
\begin{equation}
Y_{lm}(\theta, \phi)=\begin{cases} \sqrt{\frac{2l+1}{4\pi}\frac{(l-m)!}{(l+m)!}}P_{lm}(\cos \theta)\exp(im\phi)\quad m\geq 0 \\
(-1)^m \overline{Y}_{l-m}(\theta, \phi) \quad m<0,
\end{cases}
\end{equation}
where $P_{lm}$ represents the associated Legendre polynomial with subscripts $l$ and $m$.
\end{mydef}
The subscripts $l$ and $m$ entail the pattern of the waves along the latitude and longitude, where $m$ is the longitudinal wave number and $(l-m+1)/2$ is the latitudinal wave number.
Proposition \ref{addition} gives the addition formula for the spherical harmonics.
\begin{myprop}\label{addition}
(Addition Formula) For any $\bm{x}, \bm{y} \in \mathbb{S}^2$,
\begin{equation}
\sum_{m=-l}^{l} Y_{lm}(\bm{x}) \overline{Y}_{lm}(\bm{y})=\frac{2l+1}{4\pi}P_l(\langle \bm{x}, \bm{y} \rangle),
\end{equation}
where $\langle \cdot, \cdot \rangle$ denotes the inner product on $\mathbb{R}^3$ and $P_l$ represents the $l$-th Legendre polynomial.
\end{myprop}

\section{Spherical Needlets}
The spherical harmonics are globally supported on the sphere and thus suffer from the same difficulties as the Fourier basis functions. For example, the Gibbs phenomenon may occur when a spherical function with small spatial scale features is approximated by a finite series of spherical harmonics. In this case, noticeable spurious global oscillations will be introduced and higher order of spherical harmonics are required to achieve a better approximation. 

In this paper, we shall review a new generation of spherical wavelets, called spherical needlets \citep{Narcowich-etal06, Marinucci-etal08, baldi2009, Marinucci2011}. They are not only exactly localized at a finite number of frequencies, but also decay quasi-exponentially fast away from their global maximum.

\subsection{Construction of Spherical Needlets}
The construction of the spherical needlets is based on two main ideas, which are the discretization of the sphere by an exact quadrature formula and a Littlewood-Paley decomposition. Theorem \ref{quadrature} gives the exact quadrature formula.
\begin{mythm}\label{quadrature}
Denote $\mathcal{H}_l$ as the space spanned by $\{ Y_{lm}: m=-l,\cdots,l \}$, and let $\mathcal{K}_l=\bigoplus_{k=0}^l \mathcal{H}_k$. For any $l \in \mathbb{N}$, there exist a finite subset $\mathcal{X}_l=\{ \bm{\xi}_{lk}: k=1,\cdots, n_l  \}$ of $\mathbb{S}^2$ and positive real numbers $\{ \lambda_{lk}: k=1,\cdots, n_l \}$ such that 
\begin{equation}
\int_{\mathbb{S}^2} f(\bm{x})d\bm{x} = \sum \limits_{k=1}^{n_l} \lambda_{lk} f(\bm{\xi}_{lk}),
\end{equation}
for any $f \in \mathcal{K}_l$.
Here $\bm{\xi}_{lk}$ and $\lambda_{lk}$ are called cubature points and cubature weights, respectively.
\end{mythm}
The quadrature formula discretizes the sphere into cubature points and cubature weights, based on which, we present the definition of the spherical needlets.
\begin{mydef}
For a given frequency $j \in \mathbb{N}_0$ and the corresponding cubature points $\bm{\xi}_{jk}$ and cubature weights $\lambda_{jk}$ (which are obtained by Theorem \ref{quadrature} with $l=2 \lfloor B^{j+1} \rfloor$), the spherical needlets with frequency $j$ are defined as 
\begin{dmath}\label{needlet}
\psi_{jk}(\bm{x})=\sqrt{\lambda_{jk}} \sum \limits_{l=\lceil B^{j-1} \rceil}^{\lfloor B^{j+1} \rfloor} b\left( \frac{l}{B^j} \right) \sum \limits_{m=-l}^{l} Y_{lm}(\bm{\xi}_{jk})\overline{Y}_{lm}(\bm{x})
=\sqrt{\lambda_{jk}} \sum \limits_{l=\lceil B^{j-1} \rceil}^{\lfloor B^{j+1} \rfloor} b\left( \frac{l}{B^j} \right) \frac{2l+1}{4\pi} P_l(\langle \bm{\xi}_{jk}, \bm{x} \rangle) \quad \mbox{(by Proposition \ref{addition})},
\end{dmath}
where $\bm{x} \in \mathbb{S}^2$, $B>1$ is a parameter and $b(\cdot)$ is a window function satisfying 
\begin{enumerate}
\item $b(\cdot)>0$ in $(1/B, B)$, and it equals zero otherwise.
\item For any $\eta \geq1$,
\begin{equation}
\sum \limits_{j=0}^{\infty} b^2 \left( \frac{\eta}{B^j} \right)=1.
\end{equation}
\item $b(\cdot)$ is $M$ times continuously differentiable for some $M=1,2,\cdots$ or $M=\infty$.
\end{enumerate}
\end{mydef}
Equation (\ref{needlet}) implies that the spherical needlets are real-valued functions. The window function $b(\cdot)$ plays the role of the Littlewood-Paley decomposition, i.e., it decomposes the frequency domain into several overlapping intervals $(B^{j-1}, B^{j+1}),  j=0,1,\cdots$.
Here $\bm{\xi}_{jk}$ represents the location (i.e., center) of $\psi_{jk}$, while $j$ determines to what extent $\psi_{jk}$ is spatially localized. The larger $j$, the finer $\psi_{jk}$. Varying the values of $\bm{\xi}_{jk}$ and $j$ has the same effect as the translation and dilation in a multiresolution analysis. Thus, compared with the spherical harmonics,  the spherical needlets are more suited to represent spherical functions with sharp local peaks or valleys.
\subsection{Properties of Spherical Needlets}
Compared with other spherical wavelets, the spherical needlets possess several attractive properties:
\begin{enumerate}
\item They are exactly localized in the frequency domain since the window function $b(\cdot)$ has compact support.
\item They are also localized in the spatial domain. Specifically, we have
\begin{equation}
\lvert \psi_{jk}(\bm{x})\rvert\leq \frac{c_M B^j}{[1+B^j \arccos(\langle \bm{\xi}_{jk},\bm{x} \rangle)]^M} \quad \mbox{for any } \bm{x} \in \mathbb{S}^2,
\end{equation}
where $M\in \mathbb{N}$ such that $b$ is M times continuously differentiable, and $c_M$ is a constant regardless of $j$ and $k$.
\item The spherical needlets (together with the first spherical harmonic $Y_{00}=\sqrt{1/(4\pi)}$) constitute a Parseval tight frame, i.e., for any $T \in L^2(\mathbb{S}^2)$,
\begin{equation}\label{energy}
A\lVert T \rVert_2^2 \leq \sum \limits_{j,k} \lvert \langle T, \psi_{jk} \rangle \rvert^2+a_{00}^2 \leq B\lVert T \rVert_2^2,
\end{equation}
where $A=B=1$, and 
$$\langle T, \psi_{jk} \rangle=\int_{\mathbb{S}^2} T(\bm{x})\psi_{jk}(\bm{x})d\bm{x},$$ 
which are called needlet coefficients and denoted as $\beta_{jk}$,
\begin{dmath}\label{needlet_coef}
\beta_{jk}=\int_{\mathbb{S}^2} T(\bm{x})\psi_{jk}(\bm{x})d\bm{x}
=\sqrt{\lambda_{jk}}\sum \limits_{l=0}^{\infty} b\left( \frac{l}{B^j} \right) \sum \limits_{m=-l}^{l} a_{lm} Y_{lm}(\bm{\xi}_{jk})
=\sqrt{\lambda_{jk}}\sum \limits_{l=\lceil B^{j-1} \rceil}^{\lfloor B^{j+1} \rfloor} b\left( \frac{l}{B^j} \right) \sum \limits_{m=-l}^{l} a_{lm} Y_{lm}(\bm{\xi}_{jk}),
\end{dmath}
where $a_{lm}$ are the spherical harmonic coefficients defined by equation (\ref{alm}).

Equation (\ref{energy}) entails that (1) the needlet expansion preserves the ``energy" of the function in the sense of L2-norm; (2) the dual frame coincides with the frame itself, and thus the spherical needlets have the perfect reconstruction property of orthonormal bases
\begin{equation}\label{reconstruction}
T(\bm{x})=a_{00}Y_{00}(\bm{x})+\sum \limits_{j,k} \beta_{jk}\psi_{jk}(\bm{x}).
\end{equation}
\item The spherical needlets are almost orthogonal, i.e., if  $\lvert j-j' \rvert \geq 2$,
$$\langle \psi_{jk}, \psi_{j'k'} \rangle=\int_{\mathbb{S}^2}\psi_{jk}(\bm{x}) \psi_{j'k'}(\bm{x}) d\bm{x} = 0.$$
\item The spherical needlets $\psi_{jk}$ and $\psi_{jk'}$ are asymptotically uncorrelated as the frequency increases and the distance between them remains fixed, i.e.,
$$\left\lvert \frac{\langle \psi_{jk}, \psi_{jk'} \rangle}{\lVert \psi_{jk} \rVert \lVert \psi_{jk'} \rVert} \right\rvert \leq \frac{C_M}{[1+B^j \arccos(\langle \bm{\xi}_{jk},\bm{\xi}_{jk'} \rangle)]^M},$$
where
$$\langle \psi_{jk}, \psi_{jk'} \rangle = \sqrt{\lambda_{jk} \lambda_{jk'}}\sum_l b^2\left(\frac{l}{B^j}\right)\frac{2l+1}{4\pi}P_l( \langle \bm{\xi}_{jk}, \bm{\xi}_{jk'} \rangle).$$
The proof is omitted here, and we refer to Lemma 3 in \citet{baldi2009}.
\end{enumerate}

\subsection{Implementation}
The needlet coefficients can be obtained by equation (\ref{needlet_coef}), which involves the spherical harmonic coefficients. When the sampling locations are on an equiangular grid, we can utilize the package  S2Kit \citep{Healy-etal03, Kostelec-etal04} to perform a fast spherical harmonic transform. When they are irregularly spaced, the spherical harmonic coefficients can be estimated empirically by
\begin{equation}\label{alm_irr}
\widehat{a}_{lm}=\sum \limits_{i=1}^{N}w_i T(\bm{x}_i) \overline{Y}_{lm}(\bm{x}_i),
\end{equation}
where $\bm{x}_i, i=1,\cdots, N$ are the sampling locations, and $w_i$ is an appropriate cubature weight determined by the surface area associated with $\bm{x}_i$. For example, we can estimate $w_i$ to be the area of the corresponding spherical polygon in the Voronoi diagram of all the sampling locations.

It is assumed that the cubature points $\bm{\xi}_{jk}$ and cubature weights $\lambda_{jk}$ are provided by the HEALPix discretization of the sphere \citep{Gorski-etal05}. Plainly speaking, the HEALPix grid discretizes the sphere into $N_{\rm pix}$ pixels with equal area, where $N_{\rm pix}=12N_{\rm side}^2$ and $N_{\rm side}$ is required to be a power of two that measures the resolution of the discretization. We specify the cubature points $\bm{\xi}_{jk}$ as the center of the pixels, and 
the cubature weights as
\begin{equation}
\lambda_{jk}=\frac{4\pi}{N_{\rm pix}},
\end{equation}
which is the equal area of the pixels.

Suppose the highest frequency of the spherical harmonics that can be reliably extracted from the observations is $l_{\max}$. The maximum index $j_{\max}$ of the spherical needlets is the maximal $j$ such that $\lceil B^{j-1} \rceil \leq l_{\max}$. For each $j$ between $0$ and $j_{\max}$ (inclusively), the value of $N_{\rm side}$ is determined by the inequality $\lfloor B^{j+1} \rfloor \leq 2N_{\rm side}$ such that the quadrature formula in Theorem \ref{quadrature} holds approximately \citep{Pietrobon-etal10}. There are other ways of discretizating the sphere with an exact quadrature formula and (almost) equal cubature weights, such as GLESP \citep{Doroshkevich2005} and (symmetric) spherical t-designs \citep{Womersley-15}.

Computing the needlet coefficients by equation (\ref{needlet_coef}) is equivalent to performing an inverse spherical harmonic transform. The latter can be done in an efficient way by utilizing the distinct structure of the HEALPix grid. Specifically, the cubature points are arranged on a number of iso-latitude rings. The points on each ring share the same value of the co-latitude. Thus, the associated Legendre polynomials in the definition of the spherical harmonics only need to be evaluated once for all the points on each ring. Moreover, the grid is symmetric with respect to the equator, which reduces by half the computational cost of evaluating the associated Legendre polynomials. The inverse fast Fourier transform is also incorporated to further reduce the computational cost (see Appendix). Other computational techniques, such as parallelization and butterfly matrix compression are discussed in \citet{Hupca-etal12, Seljebotn-2012}.

\section{NeedMat: A Matlab Package for Spherical Needlets}
Currently, there are several packages available for the spherical needlets, such as NeedATool \citep{Pietrobon-etal10} and S2LET \citep{Leistedt-2013}. The former is written in Fortran and the latter is not easy to install because of its dependencies. In this section, we shall introduce a Matlab package for the spherical needlets, called \href{https://github.com/minjay/NeedMat}{NeedMat}, which is easy to install and use. In the package, the window function $b(\cdot)$ is constructed according to \citet{Marinucci-etal08}.
\subsection{Plot of Spherical Needlets}
\begin{figure}[htbp] %  figure placement: here, top, bottom, or page
   \centering
   \includegraphics[width=4in]{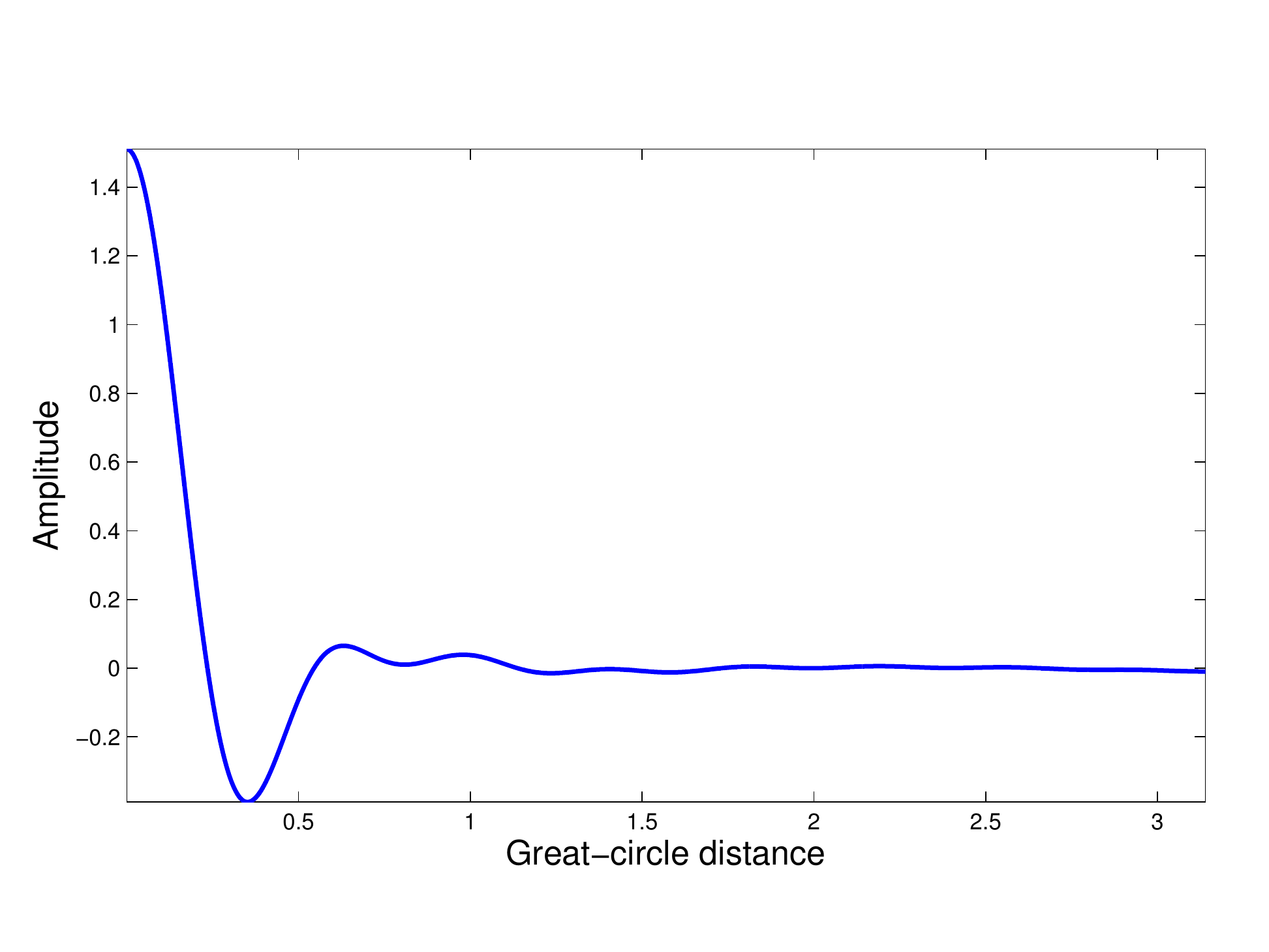} 
   \caption{Plot of the spherical needlet $\psi_{jk}$ with $j=3$ and $k=100$ as a function of the great-circle distance between $\bm{\xi}_{jk}$ and $\bm{x}$. Note that this plot remains the same for an arbitrary value of $k$.}
   \label{plot1}
\end{figure}

\begin{figure}[htbp] %  figure placement: here, top, bottom, or page
   \centering
   \includegraphics[width=3.5in]{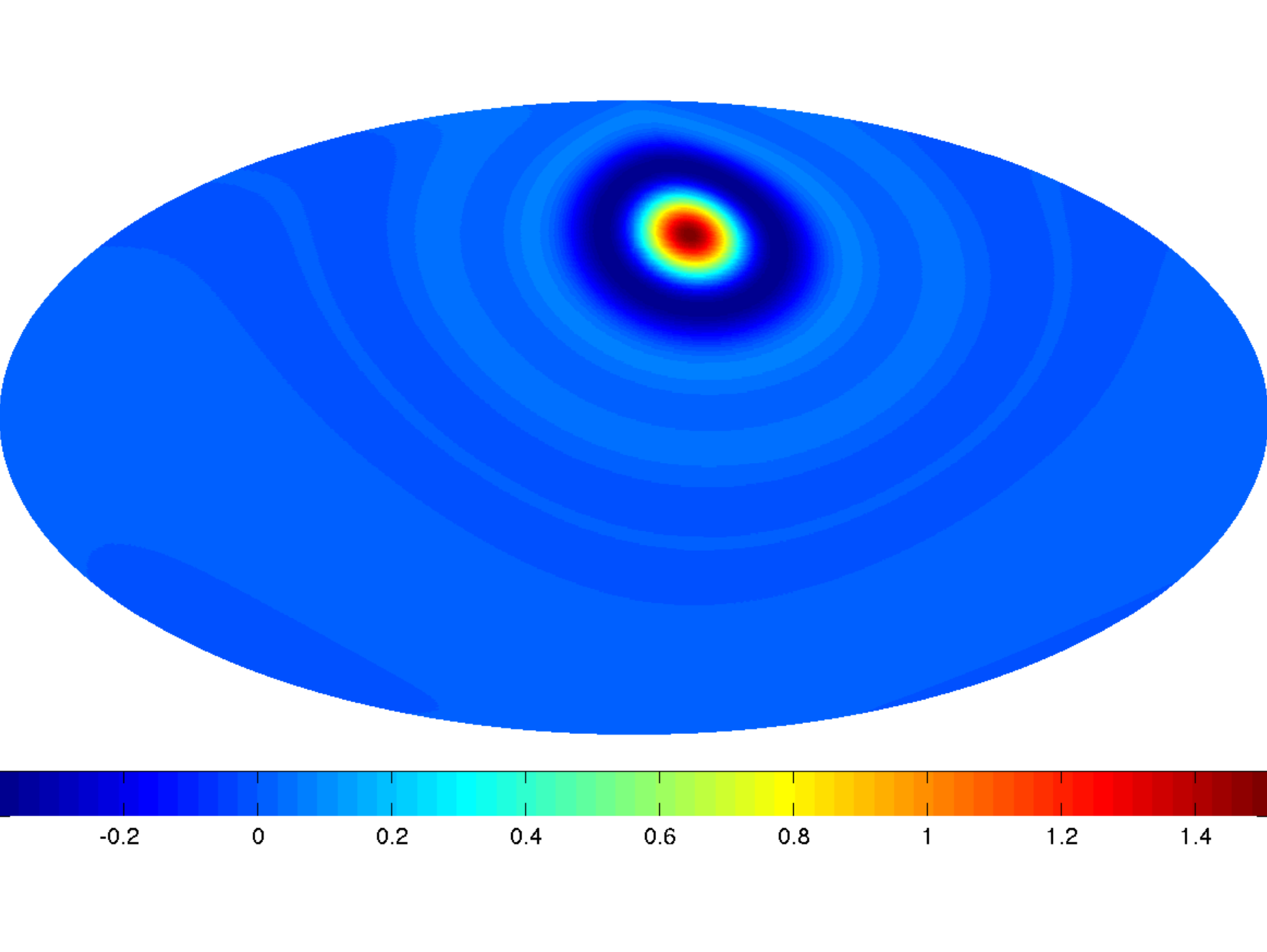} 
   \caption{Plot of the spherical needlet $\psi_{jk}$ with $j=3$ and $k=100$ on the sphere. The sphere has been projected to an ellipse by the Hammer projection.}
   \label{plot2}
\end{figure}

The spherical needlet $\psi_{jk}$ can be plotted by the function 
\begin{lstlisting}
plot_needlets(B, j, k, res);
\end{lstlisting}
where
the argument {\tt res} is a parameter controlling the resolution of the plots.
Figures \ref{plot1}-\ref{plot2} show the plots of the spherical needlet with $j=3$ and $k=100$.


\subsection{Spherical Needlet Transform}
\begin{figure}[htbp] %  figure placement: here, top, bottom, or page
   \centering
   \includegraphics[width=4in]{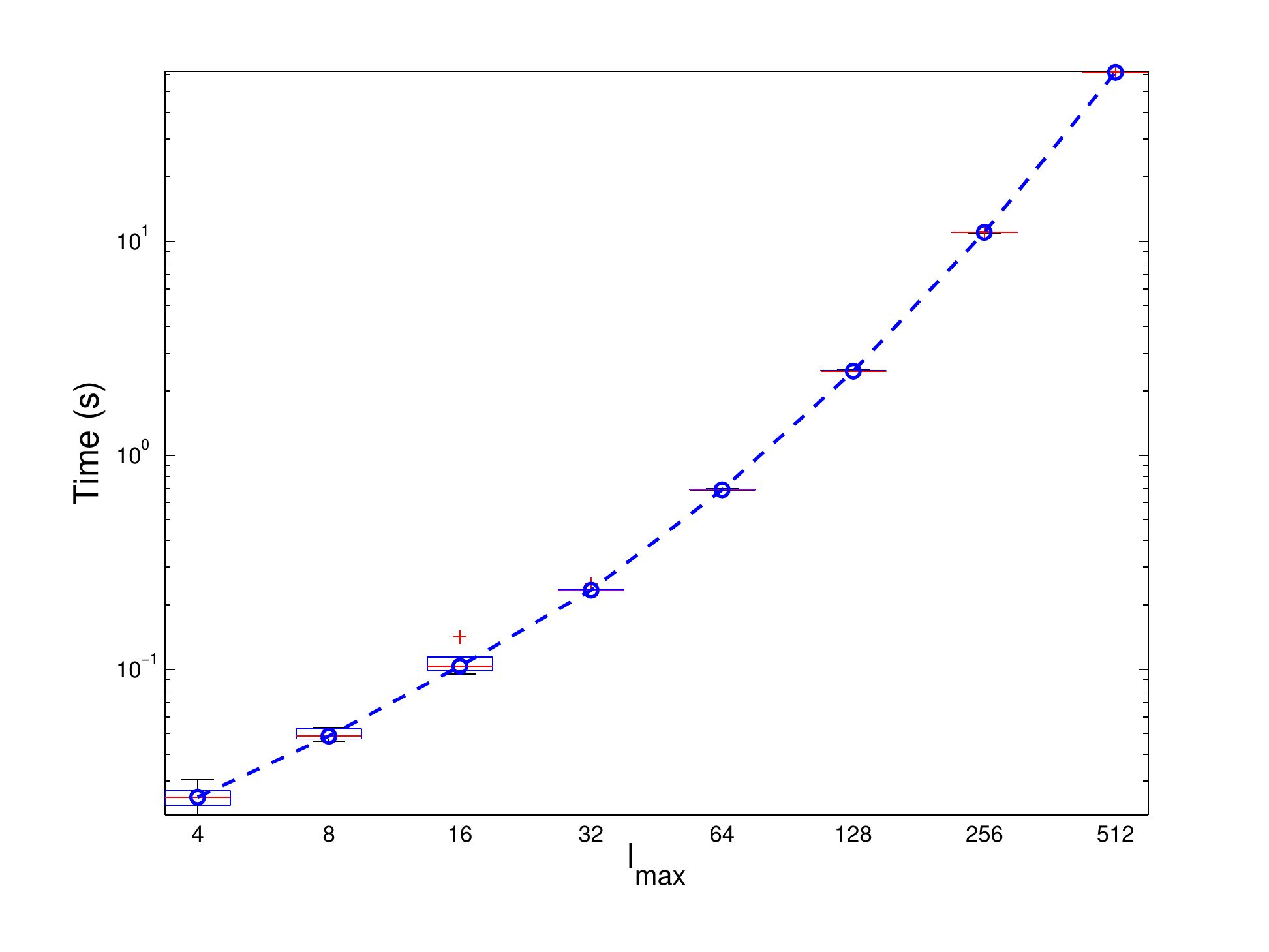} 
   \caption{Computation time of the spherical needlet transform for different $l_{\max}$.}
   \label{run_time}
\end{figure}

\begin{figure}[htbp] %  figure placement: here, top, bottom, or page
   \centering
   \includegraphics[width=3.5in]{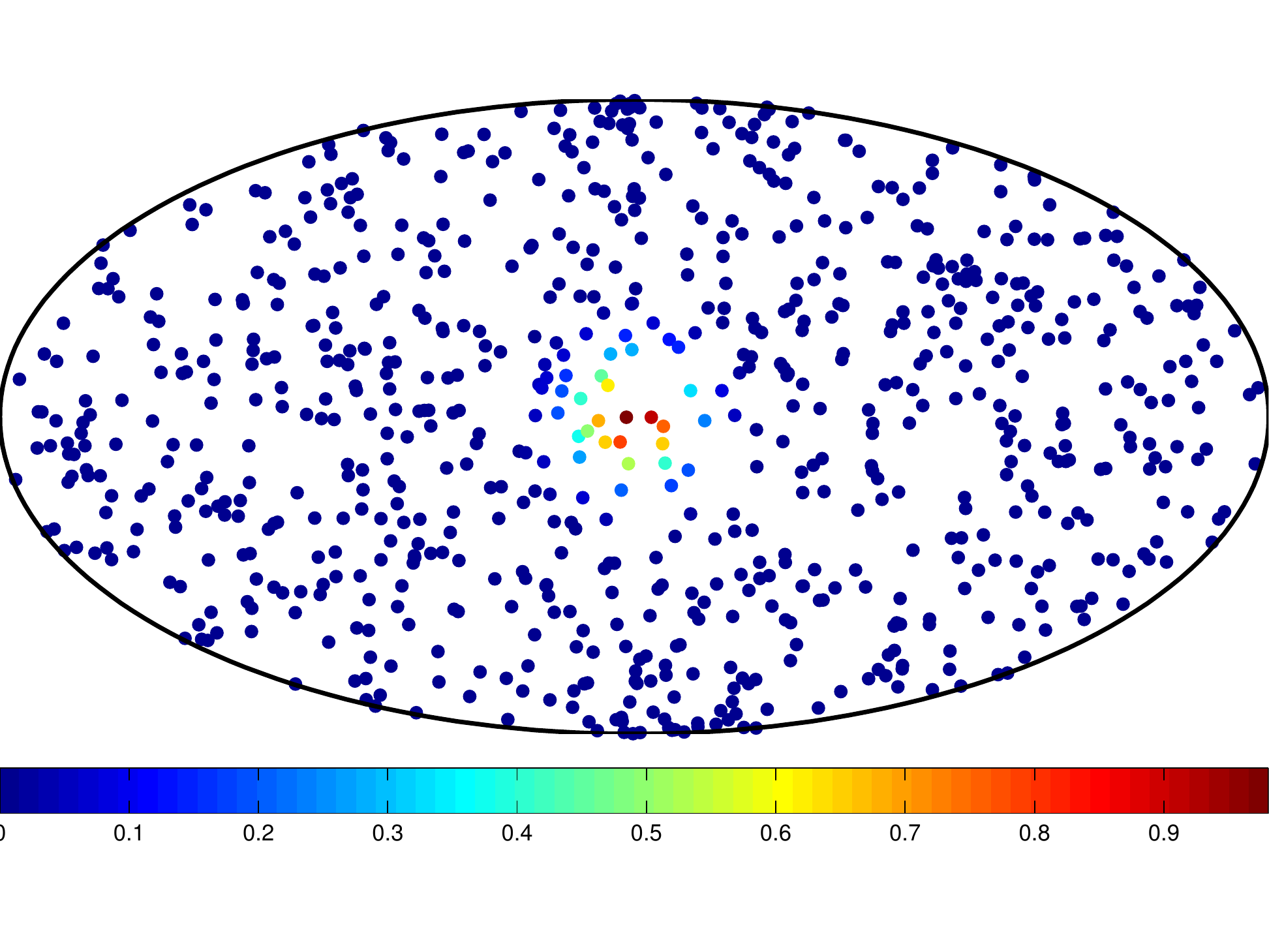} 
   \caption{Observations of $\varphi(\cdot)$ on a perturbed HEALPix grid with $768$ grid points.}
   \label{wendland}
\end{figure}

The spherical needlet transform can be performed by the function 
\begin{lstlisting}
beta = spneedlet_tran(alm, l_max, B);
\end{lstlisting}
We examine the computation time of the spherical needlet transform for different $l_{\max}$ with $a_{lm}$ randomly generated, as shown in Figure \ref{run_time}. All tests were run on a regular
laptop with a 2.4 GHz Intel Core i5 processor.
\subsection{Example: Approximation of a Wendland Radial Basis Function}
In this subsection, we present an example to demonstrate the usage of NeedMat and the superiority of the spherical needlets to the spherical harmonics in capturing small spatial scale features. The function to be approximated is
$$\varphi(\bm{x})=\varphi_0\left(\arccos (\langle \bm{\xi}_0, \bm{x} \rangle)/\rho\right),$$
where $\bm{\xi}_0$ is the center of the function, $\rho$ is a spatial scale parameter, and $\varphi_0(\cdot)$ is the Wendland radial basis function \citep{Wendland-1995} in $C^4$
$$\varphi_0(d) = 
\begin{cases}
(1-d)^6(35d^2+18d+3)/3 & \mbox{for } 0\leq d \leq 1\\
0 & \mbox{otherwise}.
\end{cases}
$$
We specify $\bm{\xi}_0=(\pi/2, \pi)$ in spherical coordinates and $\rho=\pi/4$. This specification implies that $\varphi(\cdot)$ has compact support, which is the spherical cap centered at $\bm{\xi}_0$ with radius $\pi/4$. The observations are on a perturbed HEALPix grid with $768$ grid points, as shown in Figure \ref{wendland}.  The spherical harmonic coefficients are estimated by equation (\ref{alm_irr}), where $w_i$ are determined by the Voronoi diagram. It is achieved by performing
\begin{lstlisting}
alm = spharmonic_tran_irr(theta, phi, f_wend, l_max);
\end{lstlisting}
where $l_{\max}$ is set as $16$.

\begin{figure}[htbp] %  figure placement: here, top, bottom, or page
   \centering
   \includegraphics[width=3.5in]{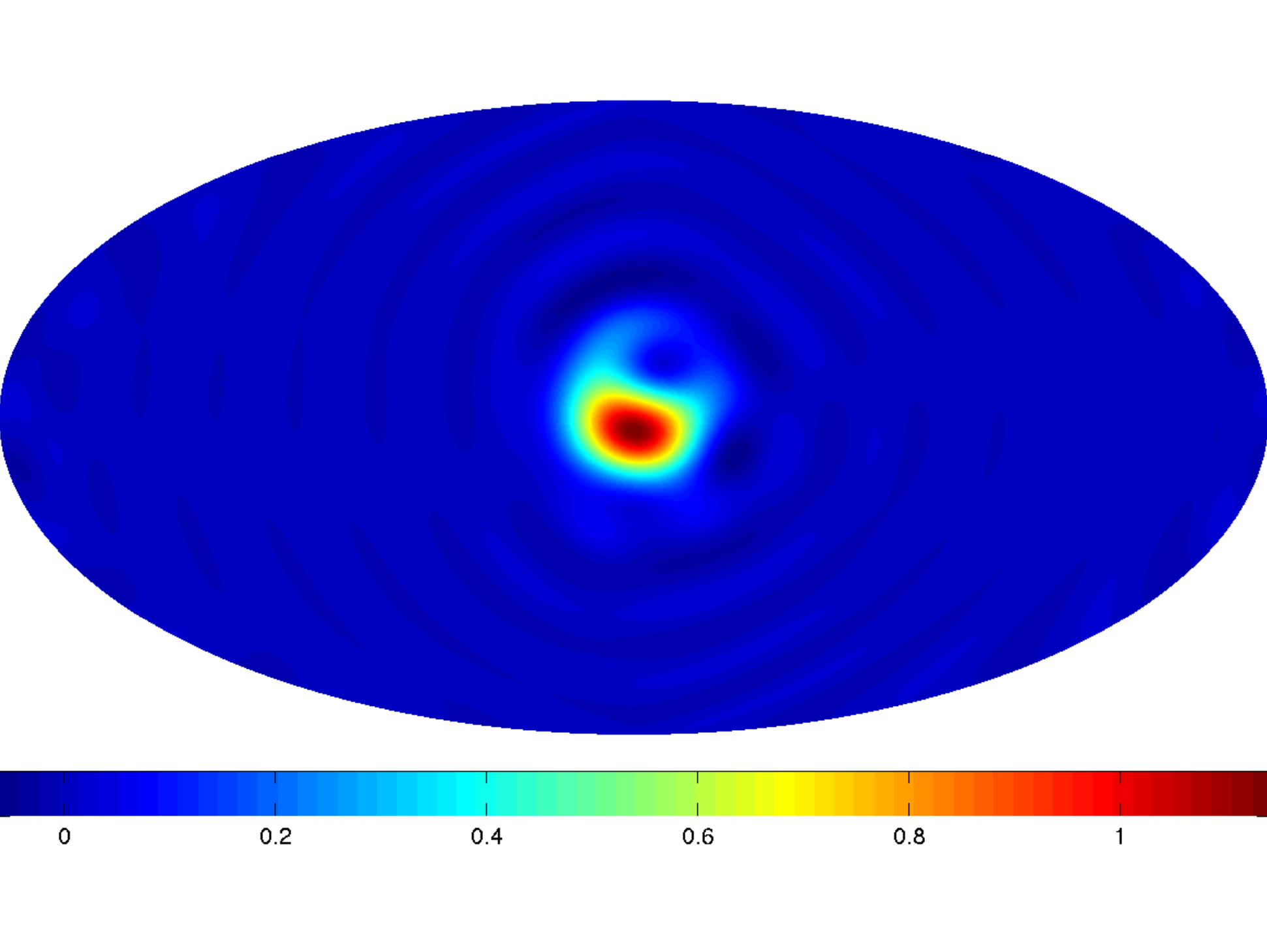} 
   \caption{Function reconstructed by the spherical harmonics with $l_{\max}=16$.}
   \label{fit1}
\end{figure}

\begin{figure}[htbp] %  figure placement: here, top, bottom, or page
   \centering
   \includegraphics[width=4in]{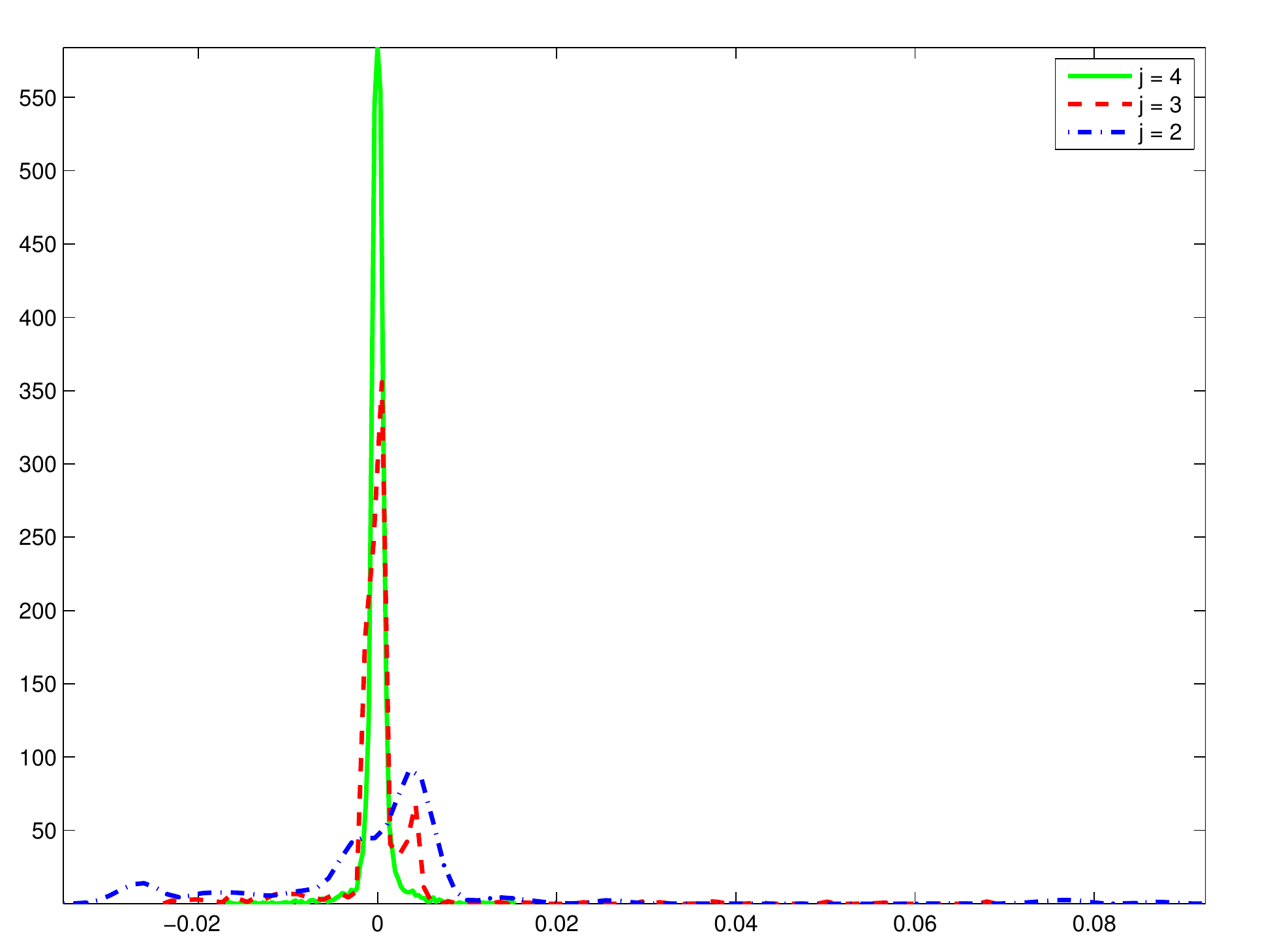} 
   \caption{Estimated probability density functions of the needlet coefficients at frequencies $j=2,3,4$.}
   \label{sparse}
\end{figure}

We reconstruct the function by the estimated spherical harmonic coefficients, as shown in Figure \ref{fit1}. The reconstructed function has noticeable spurious global oscillations, which are the artifacts induced by the spherical harmonics. Based on the estimated spherical harmonic coefficients, we obtain the needlet coefficients by equation (\ref{needlet_coef}). It is achieved by performing
\begin{lstlisting}
beta = spneedlet_tran(alm, l_max, B);
\end{lstlisting}
Figure \ref{sparse} shows the estimated probability density functions of the needlet coefficients at frequencies $j=2,3,4$. The spiky probability density functions entail that the spherical needlets can parsimoniously represent this spatially localized function, which is one of the advantages of the spherical needlets over the spherical harmonics. Theoretically, without applying any shrinking operator to the needlet coefficients, the function reconstructed by the spherical needlets should be exactly the same as that reconstructed by the spherical harmonics because of the exact transformation between them. In practice, however, the needlet coefficients are shrunk to zero to achieve a sparse representation and remove the artifacts in the function reconstructed by the spherical harmonics.

\begin{figure}[htbp] %  figure placement: here, top, bottom, or page
   \centering
   \includegraphics[width=3.5in]{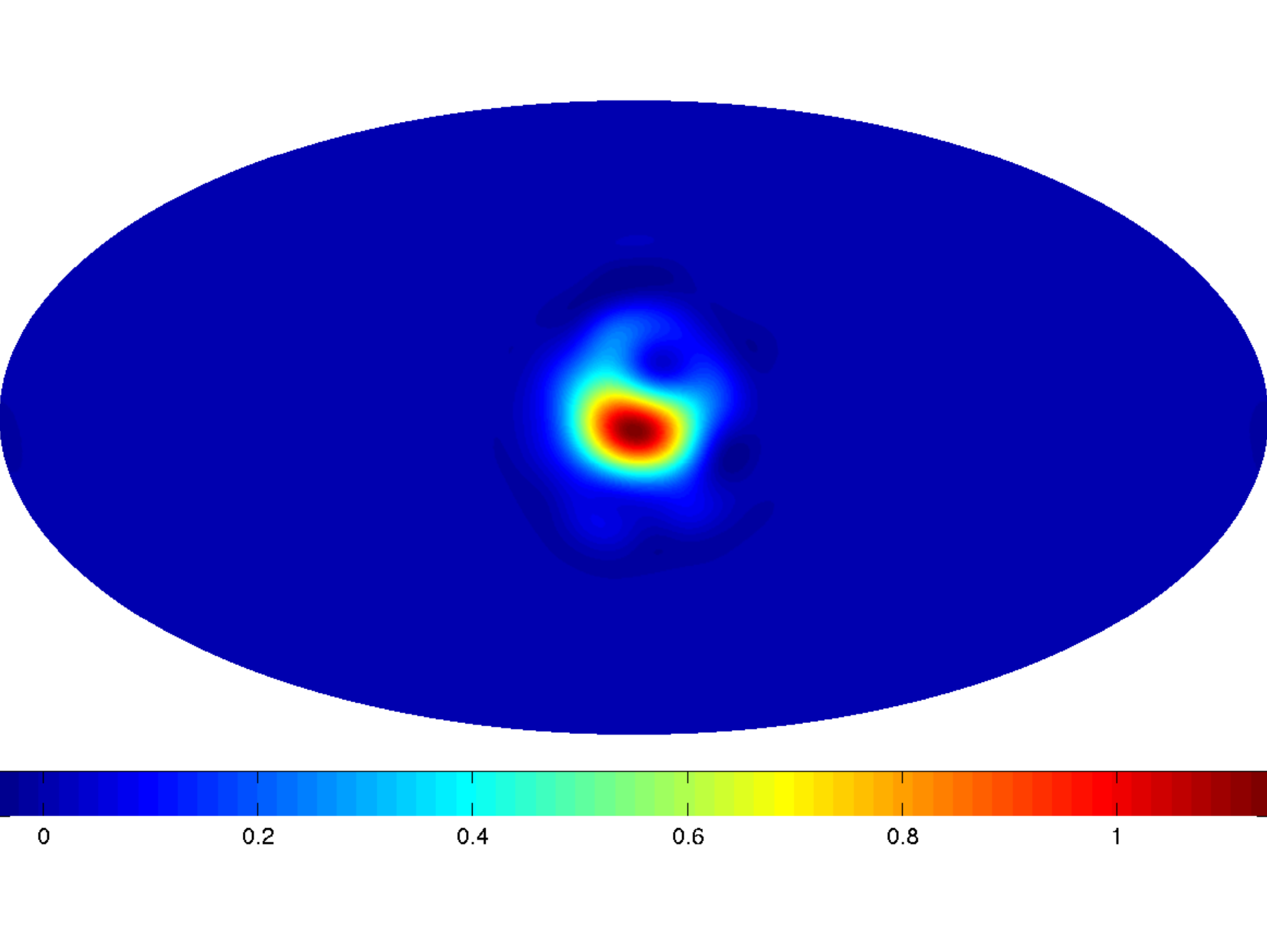} 
   \caption{Function reconstructed by the spherical needlets after hard thresholding.}
   \label{fit2}
\end{figure}

For illustration, we apply a naive percentage-based hard thresholding method to the needlet coefficients at the highest frequency $j=4$. The 95 percent of the needlet coefficients with smaller magnitudes are shrunk to zero. Figure \ref{fit2} shows the function reconstructed by the spherical needlets after thresholding. We can see that there is no such spurious global oscillation in the reconstructed function because the corresponding needlet coefficients  have been shrunk to zero. The values of these needlet coefficients are essentially artifacts that induced by the spherical harmonics. Moreover, thanks to the spatial localization of the spherical needlets, the thresholding of the needlet coefficients does not affect the reconstruction of the function within its support. More sophisticated methods are discussed in \citet{Scott-2011}.

\section*{Acknowledgement}
This research is partially supported by NSF grants AGS-1025089, PLR-1443703 and DMS-1407530. I am grateful to my advisors, Thomas C.M. Lee, Debashis Paul and Tomoko Matsuo, for their kindness, guidance, and encouragement.

\section*{Appendix: Fast Inverse Spherical Harmonic Transform}
We are interested in the following inverse spherical harmonic transform
$$\sum \limits_{l=l_{\rm st}}^{l_{\rm en}}\sum \limits_{m=-l}^l a_{lm} Y_{lm}(\bm{\xi}_{jk}),$$
where $\bm{\xi}_{jk}$ is one of the points on the HEALPix grid. Suppose $\bm{\xi}_{jk}$ is the $p$-th point ($0\leq p \leq n_r-1$) on the $r$-th ring ($1\leq r \leq N_{\rm ring}$). In spherical coordinates, 
$\bm{\xi}_{jk}=(\theta_r, \phi_{rp})$, where $\phi_{rp}=\phi_{r0}+2\pi p/n_r$.
Using the fact that $a_{l-m}=(-1)^m \overline{a}_{lm}$ and $Y_{l-m}=(-1)^m \overline{Y}_{lm}$, we have
$$
\sum \limits_{l=l_{\rm st}}^{l_{\rm en}}\sum \limits_{m=-l}^l a_{lm} Y_{lm}(\bm{\xi}_{jk})
=\RN{1}+\RN{2}+\overline{\RN{2}}, 
$$
where 
$$\RN{1}=\sum \limits_{l=l_{\rm st}}^{l_{\rm en}} a_{l0}Y_{l0} (\bm{\xi}_{jk}),$$
and
$$\RN{2}=\sum \limits_{l=l_{\rm st}}^{l_{\rm en}}\sum \limits_{m=1}^l a_{lm} Y_{lm}(\bm{\xi}_{jk}).$$
Define $\widetilde{P}_{lm}$ as the normalized associated Legendre polynomial with subscripts $l$ and $m$,
$$\widetilde{P}_{lm}(x)=\sqrt{\frac{2l+1}{4\pi}\frac{(l-m)!}{(l+m)!}}P_{lm}(x).$$
Then
$$\RN{1}=\sum \limits_{l=l_{\rm st}}^{l_{\rm en}} a_{l0} \widetilde{P}_{l0}(\cos \theta_r),$$
and
$$\RN{2}=\sum \limits_{l=l_{\rm st}}^{l_{\rm en}} \sum \limits_{m=1}^l a_{lm} \widetilde{P}_{lm}(\cos \theta_r) \exp(im \phi_{rp}).$$
Interchanging the order of summation in $\RN{2}$, we have
\begin{dmath*}
\RN{2}=\sum \limits_{m=1}^{l_{\rm en}} \left[\sum \limits_{l= \max \{ m, l_{\rm st} \}}^{l_{\rm en}}a_{lm} \widetilde{P}_{lm}(\cos \theta_r)\right] \exp(im \phi_{rp})=\sum \limits_{m=1}^{l_{\rm en}} q_{mr}  \exp(im \phi_{rp}).
\end{dmath*}
Let $m=n_rs+t$. Then
\begin{dmath*}
\RN{2}=\sum \limits_{t=0}^{n_r-1} \left\{ \sum_s q_{n_rs+t, r} \exp \left[i(n_rs+t)\phi_{r0}\right] \right\}w^{-pt}
=\sum \limits_{t=0}^{n_r-1} \tau_{tr} w^{-pt},
\end{dmath*}
where $w=\exp (-2\pi i/n_r)$.
Thus, $\RN{2}$ can be obtained by the inverse fast Fourier transform.
\bibliography{references}
\bibliographystyle{rss}

\end{document}